\documentclass[10pt,twocolumn,letterpaper]{article}

\usepackage{cvpr}
\usepackage{times}
\usepackage{epsfig}
\usepackage{graphicx}
\usepackage{amsmath}
\usepackage{amssymb}
\usepackage{subcaption}
\usepackage{fixltx2e}
\usepackage{multirow}
\usepackage{color}

\usepackage[pagebackref=true,breaklinks=true,letterpaper=true,colorlinks,bookmarks=false]{hyperref}

\cvprfinalcopy 

\def\cvprPaperID{156} 

\ifcvprfinal\fi
\begin{document}

\title{Deep Learning for Identifying Metastatic Breast Cancer}

\author{Dayong Wang\hspace*{5mm}Aditya Khosla$^\star$\hspace*{5mm}Rishab Gargeya\hspace*{5mm}Humayun Irshad\hspace*{5mm}Andrew H Beck\\
Beth Israel Deaconess Medical Center, Harvard Medical School\\
$^\star$CSAIL, Massachusetts Institute of Technology\\
{\tt\small \{dwang5,hirshad,abeck2\}@bidmc.harvard.edu}\hspace*{4mm}{\tt\small khosla@csail.mit.edu}\\
{\tt\small rishab.gargeya@gmail.com}
}

\maketitle


%

%
%
\begin{abstract}
The International Symposium on Biomedical Imaging (ISBI) held a grand challenge to evaluate computational systems for the automated detection of metastatic breast cancer in whole slide images of sentinel lymph node biopsies. Our team won both competitions in the grand challenge, obtaining an area under the receiver operating curve (AUC) of 0.925 for the task of whole slide image classification and a score of 0.7051 for the tumor localization task.
A pathologist independently reviewed the same images, obtaining a whole slide image classification AUC of 0.966 and a tumor localization score of 0.733. Combining our deep learning system's predictions with the human pathologist's diagnoses increased the pathologist's AUC to 0.995, representing an approximately 85 percent reduction in human error rate. These results demonstrate the power of using deep learning to produce significant improvements in the accuracy of pathological diagnoses. 

\end{abstract}
%
%
\section{Introduction}
The medical specialty of pathology is tasked with providing definitive disease diagnoses to guide patient treatment and management decisions \cite{cotran1999robbins}. Standardized, accurate and reproducible pathological diagnoses are essential for advancing precision medicine. Since the mid-19$^{\text{th}}$ century, the primary tool used by pathologists to make diagnoses has been the microscope \cite{ackerknecht1953rudolf}. Limitations of the qualitative visual analysis of microscopic images includes lack of standardization, diagnostic errors, and the significant cognitive load required to manually evaluate millions of cells across hundreds of slides in a typical pathologist's workday ~\cite{nakhleh2006error,raab2005clinical,elmore2015diagnostic}. Consequently, over the past several decades there has been increasing interest in developing computational methods to assist in the analysis of microscopic  images in pathology ~\cite{gurcan2009histopathological,ghaznavi2013digital}.

From October 2015 to April 2016, the International Symposium on Biomedical Imaging (ISBI) held the Camelyon Grand Challenge 2016 (Camelyon16) to identify top-performing computational image analysis systems for the task of automatically detecting metastatic breast cancer in  digital whole slide images (WSIs) of sentinel lymph node biopsies\footnote{\href{http://camelyon16.grand-challenge.org/}{http://camelyon16.grand-challenge.org/}}. The evaluation of breast sentinel lymph nodes is an important component of the American Joint Committee on Cancer's TNM breast cancer staging system, in which patients with a sentinel lymph node positive for metastatic cancer will receive a higher pathologic TNM stage than patients negative for sentinel lymph node metastasis~\cite{edge2010american}, frequently resulting in more aggressive clinical management, including axillary lymph node dissection~\cite{lyman2005american,lyman2014sentinel}.

The manual pathological review of sentinel lymph nodes is time-consuming and laborious, particularly in cases in which the lymph nodes are negative for cancer or contain only small foci of metastatic cancer. Many centers have implemented testing of sentinel lymph nodes with immunohistochemistry for pancytokeratins~\cite{czerniecki1999immunohistochemistry}, which are proteins expressed on breast cancer cells and not normally present in lymph nodes, to improve the sensitivity of cancer metastasis detection. However, limitations of pancytokeratin immunohiostochemistry testing of sentinel lymph nodes include:  increased cost, increased time for slide preparation, and increased number of slides required for pathological review. Further, even with immunohistochemistry-stained slides, the identification of small cancer metastases can be tedious and inaccurate. 

Computer-assisted image analysis systems have been developed to aid in the detection of small metastatic foci from pancytokeratin-stained immunohistochemistry slides of sentinel lymph nodes~\cite{weaver2003comparison}; however, these systems are not used clinically. Thus, the development of effective and cost efficient methods for sentinel lymph node evaluation remains an active area of research~\cite{jaffer2014evolution}, as there would be value to a high-performing system that could increase accuracy and reduce cognitive load at low cost. 

Here, we present a deep learning-based approach for the identification of cancer metastases from whole slide images of breast sentinel lymph nodes. Our approach uses millions of training patches to train a deep convolutional neural network to make patch-level predictions to discriminate tumor-patches from normal-patches. We then aggregate the patch-level predictions to create tumor probability heatmaps and perform post-processing over these heatmaps to make predictions for the slide-based classification task and the tumor-localization task. Our system won both competitions at the Camelyon Grand Challenge 2016, with performance approaching human level accuracy. Finally, combining the predictions of our deep learning system with a pathologist's interpretations produced a significant reduction in the pathologist's error rate.

\section{Dataset and Evaluation Metrics}
In this section, we describe the Camelyon16 dataset provided by the organizers of the competition and the evaluation metrics used to rank the participants.

\subsection{Camelyon16 Dataset}
The Camelyon16 dataset consists of a total of 400 whole slide images (WSIs) split into 270 for training and 130 for testing. Both splits contain samples from two institutions (Radbound UMC and UMC Utrecht) with specific details provided in Table~\ref{tab:dataset-split}. 
\begin{table}[htbp]
\centering
\caption{Number of slides in the Camelyon16 dataset.}
\label{tab:dataset-split}
\begin{tabular}{|c|c|c|c|}
\hline 
\multirow{2}{*}{Institution}  & Train & Train & \multirow{2}{*}{Test} \\
 & cancer & normal & \\
\hline
Radboud UMC & 90 & 70 & 80 \\
UMC Utrecht & 70 & 40 & 50 \\
\hline
Total & 160 & 110 & 130 \\
\hline
\end{tabular}
\end{table}

The ground truth data for the training slides consists of a pathologist's delineation of regions of metastatic cancer on WSIs of sentinel lymph nodes. The data was provided in two formats: XML files containing vertices of the annotated contours of the locations of cancer metastases and WSI binary masks indicating the location of the cancer metastasis. 

\subsection{Evaluation Metrics}
Submissions to the competition were evaluated on the following two metrics:
\begin{itemize}
\item \textbf{Slide-based Evaluation:}  For this metric, teams were judged on performance at discriminating between slides containing metastasis and normal slides. Competition participants submitted a probability for each test slide indicating its predicted likelihood of containing cancer. The competition organizers measured the participant performance using the area under the receiver operator (AUC) score.  
\item \textbf{Lesion-based Evaluation:} For this metric, participants submitted a probability and a corresponding $(x,y)$ location for each predicted cancer lesion within the WSI. The competition organizers measured participant performance as the average sensitivity for detecting all true cancer lesions in a WSI across 6 false positive rates: $\frac{1}{4}$, $\frac{1}{2}$, 1, 2, 4, and 8 false positives per WSI.
\end{itemize}

\section{Method}
In this section, we describe our approach to cancer metastasis detection. 

\subsection{Image Pre-processing}
\label{sec:preprocessing}
\begin{figure}[htbp]
	\centering
	\includegraphics[width=1.6in, height=1.6in]{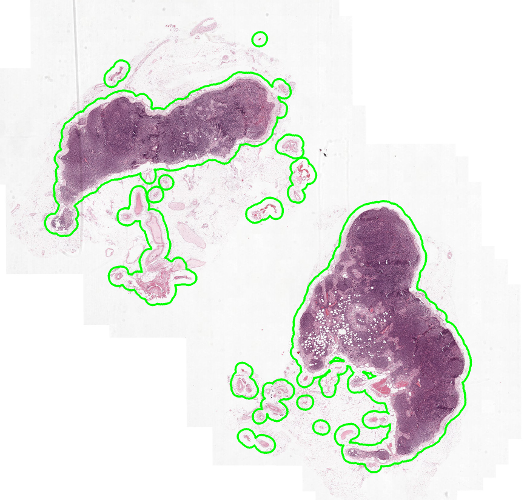} 
	\includegraphics[width=1.6in, height=1.6in]{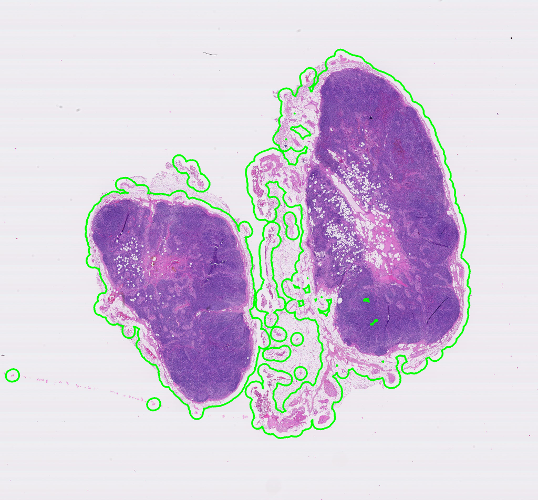} \\
	\makebox[1.6in]{(a)}
	\makebox[1.6in]{(b)}
	\caption{Visualization of tissue region detection during image pre-processing (described in Section~\ref{sec:preprocessing}). Detected tissue regions are highlighted with the green curves.}\label{fig:tissue_detection}
\end{figure}

To reduce computation time and to focus our analysis on regions of the slide most likely to contain cancer metastasis, we first identify tissue within the WSI and exclude background white space. To achieve this, we adopt a threshold based segmentation method to automatically detect the background region. In particular, we first transfer the original image from the RGB color space to the HSV color space, then the optimal threshold values in each channel are computed using the Otsu algorithm~\cite{otsu},  and the final mask images are generated by combining the masks from H and S channels. The detection results are visualized in Fig.~\ref{fig:tissue_detection}, where the tissue regions are highlighted using green curves. According to the detection results, the average percentage of background region per WSI is approximately $82\%$.

\subsection{Cancer Metastasis Detection Framework}
\begin{figure*}[htbp]
	\centering
	\includegraphics[width=\linewidth]{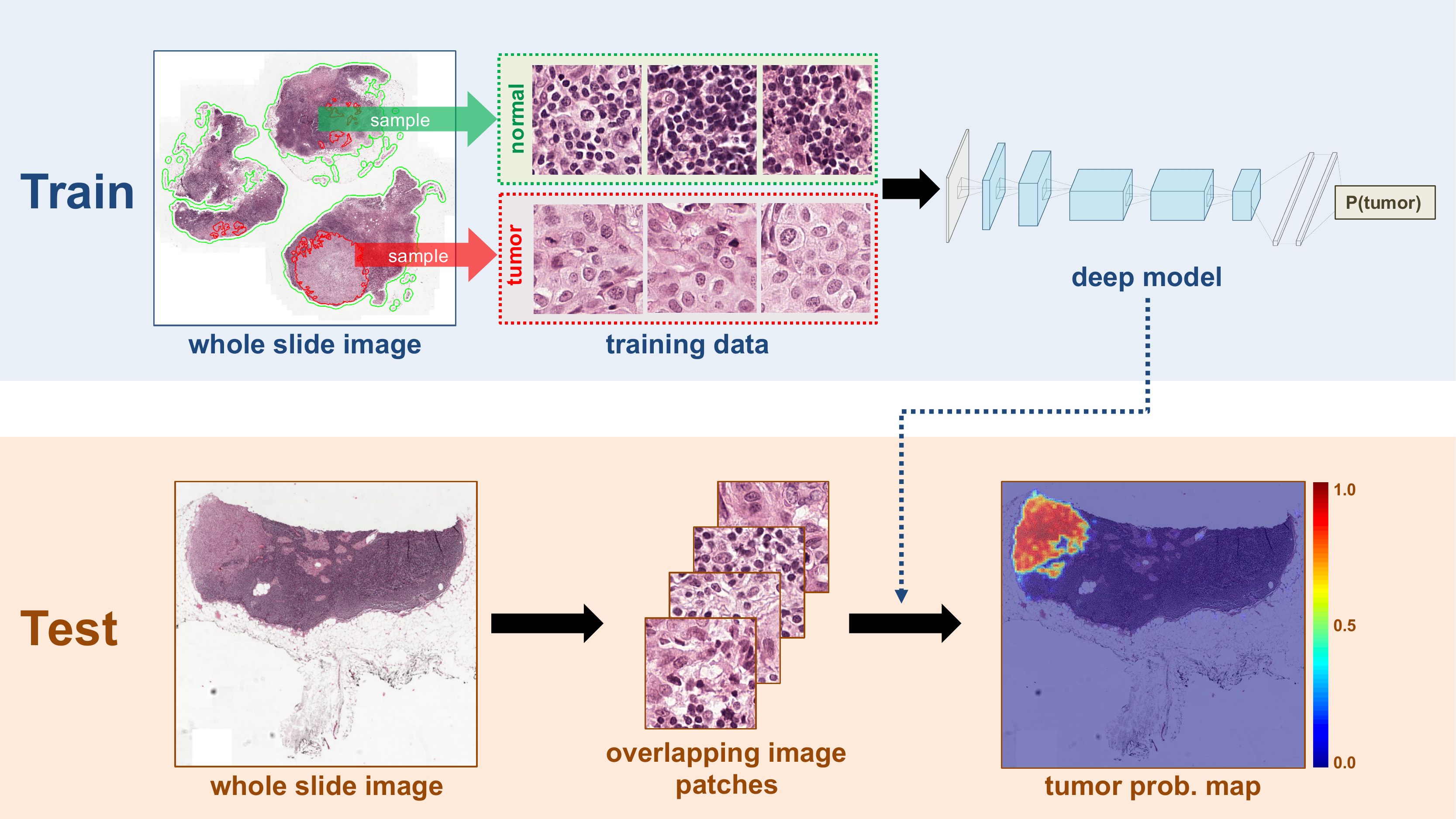} 
	\caption{The framework of cancer metastases detection.} \label{fig:framework}
\end{figure*}
Our cancer metastasis detection framework consists of a \emph{patch-based classification} stage and a \emph{heatmap-based post-processing} stage, as depicted in Fig.~\ref{fig:framework}. 

During model training, the \textbf{patch-based classification} stage takes as input whole slide images and the ground truth image annotation, indicating the locations of regions of each WSI containing metastatic cancer.  We randomly extract millions of small \emph{positive} and \emph{negative} patches from the set of training WSIs. If the small patch is located in a tumor region, it is a tumor / positive patch and labeled with $1$, otherwise, it is a normal / negative patch and labeled with $0$. Following selection of positive and negative training examples, we train a supervised classification model to discriminate between these two classes of patches, and we embed all the prediction results into a heatmap image. In the \textbf{heatmap-based post-processing} stage, we use the tumor probability heatmap to compute the slide-based evaluation and lesion-based evaluation scores for each WSI. 

\subsection{Patch-based Classification Stage}
During training, this stage uses as input 256x256 pixel patches from positive and negative regions of the WSIs and trains a classification model to discriminate between the positive and negative patches. We evaluated the performance of four well-known deep learning network architectures for this classification task: GoogLeNet~\cite{googlenet}, AlexNet~\cite{alexnet}, VGG16~\cite{vgg16} and a face orientated deep network~\cite{face}, as shown in Table~\ref{tab:deep-model-performance}. The two deeper networks (GoogLeNet and VGG16) achieved the best patch-based classification performance. In our framework, we adopt GoogLeNet as our deep network structure since it is generally faster and more stable than VGG16. The network structure of GoogLeNet consists of $27$ layers in total and more than $6$ million parameters. 

\begin{table}[htbp]
\centering
\caption{Evaluation of Various Deep Models}
\label{tab:deep-model-performance}
\begin{tabular}{|l|c|}
\hline
          & Patch classification accuracy \\ \hline
GoogLeNet~\cite{googlenet} & 98.4\%                           \\ \hline
AlexNet~\cite{alexnet}   & 92.1\%                           \\ \hline
VGG16~\cite{vgg16}     & 97.9\%                           \\ \hline
FaceNet~\cite{face}   & 96.8\%                           \\ \hline
\end{tabular}
\end{table}

In our experiments, we evaluated a range of magnification levels, including $40\times$, $20\times$ and $10\times$, and we obtained the best performance with $40\times$ magnification. We used only the $40\times$ magnification in the experimental results reported for the Camelyon competition.

After generating tumor-probability heatmaps using GoogLeNet across the entire training dataset, we noted that a significant proportion of errors were due to false positive classification from histologic mimics of cancer. To improve model performance on these regions, we extract additional training examples from these difficult negative regions and retrain the model with a training set enriched for these \textit{hard negative}  patches. 

\begin{figure*}[htbp]
	\centering
	\includegraphics[width=6.6in]{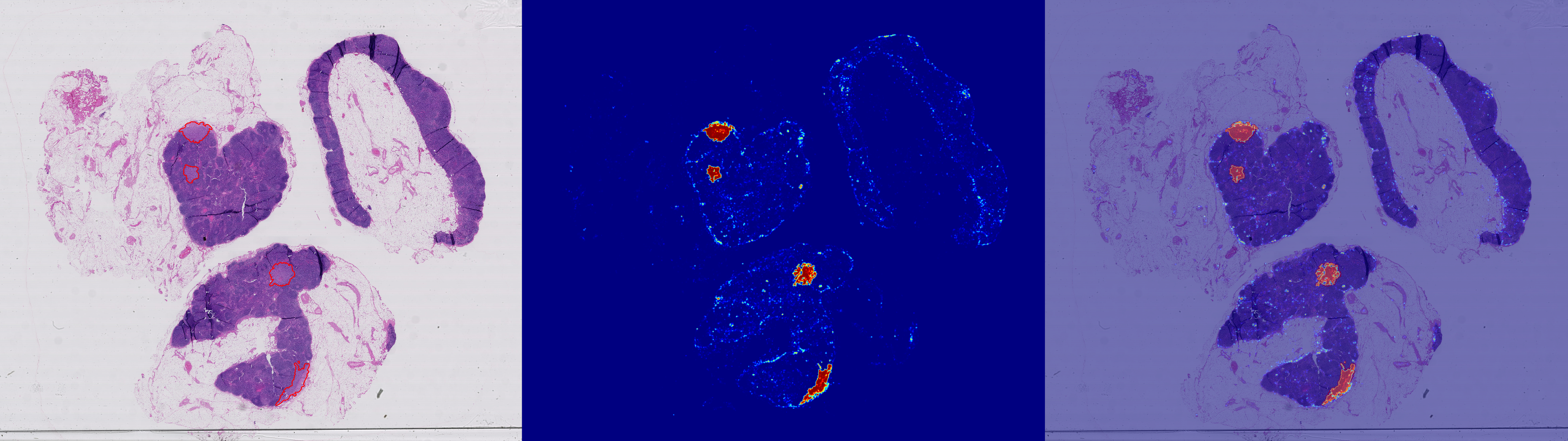} \\
	\makebox[2.2in]{(a) Tumor Slide}
	\makebox[2.2in]{(b) Heatmap}
	\makebox[2.2in]{(c) Heatmap overlaid on slide} 
	\caption{Visualization of tumor region detection.} \label{fig:results}
\end{figure*}

We present one of our results in Fig.~\ref{fig:results}. Given a whole slide image (Fig.~\ref{fig:results} (a)) and a deep learning based patch classification model, we generate the corresponding tumor region heatmap (Fig.~\ref{fig:results} (b)), which highlights the tumor area. 

\subsection{Post-processing of tumor heatmaps to compute slide-based and lesion-based probabilities}
After completion of the patch-based classification stage, we generate a tumor probability heatmap for each WSI. On these heatmaps, each pixel contains a value between 0 and 1, indicating the probability that the pixel contains tumor. 
We now perform post-processing to compute slide-based and lesion-based scores for each heatmap. 

\subsubsection{Slide-based Classification}\label{sec:wsi-level}
For the slide-based classification task, the post-processing takes as input a heatmap for each WSI and produces as output a single probability of tumor for the entire WSI. Given a heatmap, we extract 28 geometrical and morphological features from each heatmap, including the percentage of tumor region over the whole tissue region, the area ratio between tumor region and the minimum surrounding convex region, the average prediction values, and the longest axis of the tumor region. We compute these features over tumor probability heatmaps across all training cases, and we build a random forest classifier to discriminate the WSIs with metastases from the negative WSIs. On the test cases, our slide-based classification method achieved an AUC of 0.925, making it the top-performing system for the slide-based classification task in the Camelyon grand challenge. 

\subsubsection{Lesion-based Detection}\label{sec:lesion-level}
For the lesion-based detection post-processing, we aim to identify all cancer lesions within each WSI with few false positives. To achieve this, we first train a deep model (D-I) using our initial training dataset described above. We then train a second deep model (D-II) with a training set that is enriched for tumor-adjacent negative regions. This model (D-II) produces fewer false positives than D-I but has reduced sensitivity. In our framework, we first threshold the heatmap produced from D-I at 0.90, which creates a binary heatmap. We then identify connected components within the tumor binary mask, and we use the central point as the tumor location for each connected component. To estimate the probability of tumor at each of these $(x,y)$ locations, we take the average of the tumor probability predictions generated by D-I and D-II across each connected component. The scoring metric for Camelyon16 was defined as the average sensitivity at 6 predefined false positive rates: 1/4, 1/2, 1, 2, 4, and 8 FPs per whole slide image. Our system achieved a score of 0.7051, which was the highest score in the competition and was 22 percent higher than the second-ranking score (0.5761). 

\section{Experimental Results}
\subsection{Evaluation Results from Camelyon16}
In this section, we briefly present the evaluation results generated by the Camelyon16 organizers, which is also available on the website~\footnote{\href{http://camelyon16.grand-challenge.org/results/}{http://camelyon16.grand-challenge.org/results/}}.

There are two kinds evaluation in Camelyon16: \emph{Slide-based Evaluation} and \emph{Lesion-based Evaluation}. We won both of these two challenging tasks.

\begin{figure}[htbp]
	\centering
	\includegraphics[width=3.5in]{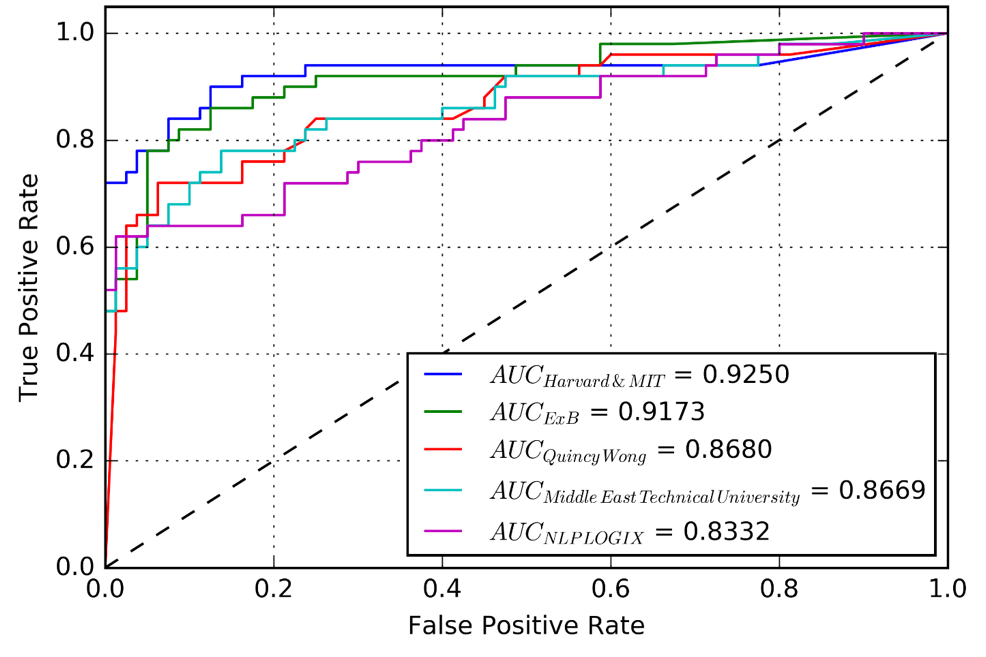} \\
	\caption{Receiver Operating Characteristic (ROC) curve of Slide-based Classification} \label{fig:result-org-01}
\end{figure}

\noindent \emph{Slide-based Evaluation}: The merits of the algorithms were assessed for discriminating between slides containing metastasis and normal slides. Receiver operating characteristic (ROC) analysis at the slide level were performed and the measure used for comparing the algorithms was area under the ROC curve (AUC). Our submitted result was generated based on the algorithm in Section~\ref{sec:wsi-level}. As shown in Fig.~\ref{fig:result-org-01}, the AUC is $0.9250$. Notice that our algorithm performed much better than the second best method when the False Positive Rate (FPR) is low.

\begin{figure}[htbp]
	\centering
    \includegraphics[width=3.5in]{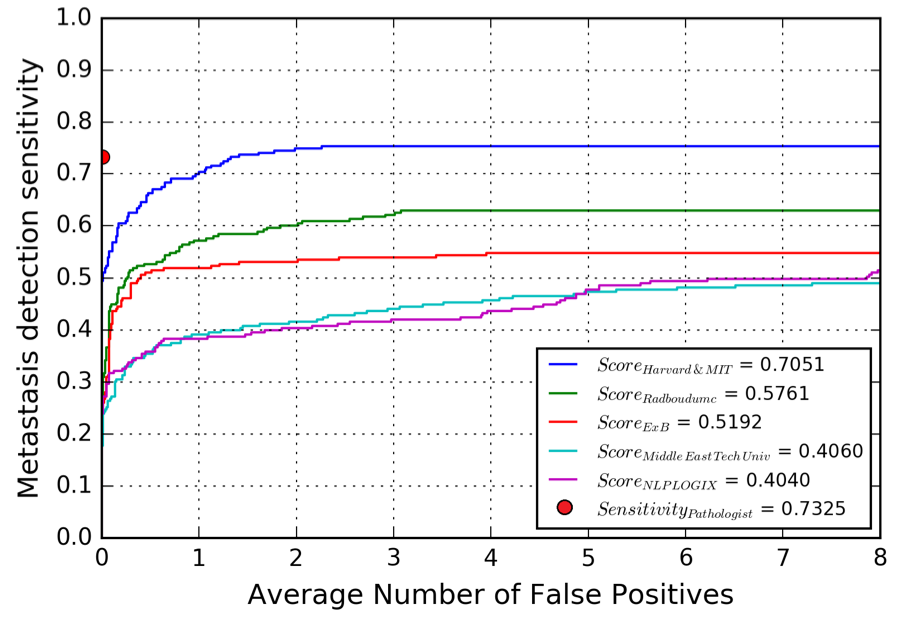} \\
	\caption{Free-Response Receiver Operating Characteristic (FROC) curve of the Lesion-based Detection.} \label{fig:result-org-02}
\end{figure}

\noindent \emph{Lesion-based Evaluation}: For the lesion-based evaluation, free-response receiver operating characteristic (FROC) curve were used. The FROC curve is defined as the plot of sensitivity versus the average number of false-positives per image. Our submitted result was generated based on the algorithm in Section~\ref{sec:lesion-level}. As shown in Fig.~\ref{fig:result-org-02}, we can make two observations: first, the pathologist did not make any false positive predictions; second, when the average number of false positives is larger than $2$, which indicates that there will be two false positive alert in each slide on average, our performance (in terms of sensitivity) even outperformed the pathologist. 

\subsection{Combining  Deep Learning System with a Human Pathologist}
To evaluate the top-ranking deep learning systems against a human pathologist, the Camelyon16 organizers had a pathologist examine the test images used in the competition. For the slide-based classification task,the human pathologist achieved an AUC of 0.9664, reflecting a 3.4 percent error rate. When the predictions of our deep learning system were combined with the predictions of the human pathologist, the AUC was raised to 0.9948 reflecting a drop in the error rate to 0.52 percent. 

\section{Discussion}
Here we present a deep learning-based system for the automated detection of metastatic cancer from whole slide images of sentinel lymph nodes. Key aspects of our system include: enrichment of the training set with patches from regions of normal lymph node that the system was initially mis-classifying as cancer; use of a state-of-the-art deep learning model architecture; and careful design of post-processing methods for the slide-based classification and lesion-based detection tasks. 

Historically, approaches to histopathological image analysis in digital pathology have focused primarily on low-level image analysis tasks (e.g., color normalization, nuclear segmentation, and feature extraction), followed by construction of classification models using classical machine learning methods, including: regression, support vector machines, and random forests. Typically, these algorithms take as input relatively small sets of image features (on the order of tens)~\cite{gurcan2009histopathological,irshad2014methods}. Building on this framework, approaches have been developed for the automated extraction of moderately high dimensional sets of image features (on the order of thousands) from histopathological images followed by the construction of relatively simple, linear classification models using methods designed for dimensionality reduction, such as sparse regression~\cite{beck2011systematic}.

Since 2012, deep learning-based approaches have consistently shown best-in-class performance in major computer vision competitions, such as the ImageNet Large Scale Visual Recognition Competition (ILSVRC)~\cite{russakovsky2015imagenet}. Deep learning-based approaches have also recently shown promise for applications in pathology. A team from the research group of Juergen Schmidhuber used a deep learning-based approach to win the ICPR 2012 and MICCAI 2013 challenges focused on algorithm development for mitotic figure detection\cite{cirecsan2013mitosis}. In contrast to the types of machine learning approaches historically used in digital pathology, in deep learning-based approaches there tend to be no discrete human-directed steps for object detection, object segmentation, and feature extraction. Instead, the deep learning algorithms take as input only the images and the image labels (e.g., 1 or 0) and learn a very high-dimensional and complex set of model parameters with supervision coming only from the image labels. 

Our winning approach in the Camelyon Grand Challenge 2016 utilized a 27-layer deep network architecture and obtained near human-level classification performance on the test data set. Importantly, the errors made by our deep learning system were not strongly correlated with the errors made by a human pathologist. Thus, although the pathologist alone is currently superior to our deep learning system alone, combining deep learning with the pathologist produced a major reduction in pathologist error rate, reducing it from over 3 percent to less than 1 percent. More generally, these results suggest that integrating deep learning-based approaches into the work-flow of the diagnostic pathologist could drive improvements in the reproducibility, accuracy and clinical value of pathological diagnoses.

\section{Acknowledgments}
We thank all the Camelyon Grand Challenge 2016 organizers with special acknowledgments to lead coordinator Babak Ehteshami Bejnordi. AK and AHB are co-founders of PathAI, Inc.

{\small
\bibliographystyle{ieee}
\bibliography{refs}

\begin{thebibliography}{10}\itemsep=-1pt

\bibitem{ackerknecht1953rudolf}
E.~H. Ackerknecht et~al.
\newblock Rudolf virchow: Doctor, statesman, anthropologist.
\newblock {\em Rudolf Virchow: Doctor, Statesman, Anthropologist.}, 1953.

\bibitem{beck2011systematic}
A.~H. Beck, A.~R. Sangoi, S.~Leung, R.~J. Marinelli, T.~O. Nielsen, M.~J.
  van~de Vijver, R.~B. West, M.~van~de Rijn, and D.~Koller.
\newblock Systematic analysis of breast cancer morphology uncovers stromal
  features associated with survival.
\newblock {\em Science translational medicine}, 3(108):108ra113--108ra113,
  2011.

\bibitem{cirecsan2013mitosis}
D.~C. Cire{\c{s}}an, A.~Giusti, L.~M. Gambardella, and J.~Schmidhuber.
\newblock Mitosis detection in breast cancer histology images with deep neural
  networks.
\newblock In {\em Medical Image Computing and Computer-Assisted
  Intervention--MICCAI 2013}, pages 411--418. Springer, 2013.

\bibitem{cotran1999robbins}
R.~S. Cotran, V.~Kumar, T.~Collins, and S.~L. Robbins.
\newblock Robbins pathologic basis of disease.
\newblock 1999.

\bibitem{czerniecki1999immunohistochemistry}
B.~J. Czerniecki, A.~M. Scheff, L.~S. Callans, F.~R. Spitz, I.~Bedrosian, E.~F.
  Conant, S.~G. Orel, J.~Berlin, C.~Helsabeck, D.~L. Fraker, et~al.
\newblock Immunohistochemistry with pancytokeratins improves the sensitivity of
  sentinel lymph node biopsy in patients with breast carcinoma.
\newblock {\em Cancer}, 85(5):1098--1103, 1999.

\bibitem{edge2010american}
S.~B. Edge and C.~C. Compton.
\newblock The american joint committee on cancer: the 7th edition of the ajcc
  cancer staging manual and the future of tnm.
\newblock {\em Annals of surgical oncology}, 17(6):1471--1474, 2010.

\bibitem{elmore2015diagnostic}
J.~G. Elmore, G.~M. Longton, P.~A. Carney, B.~M. Geller, T.~Onega, A.~N.
  Tosteson, H.~D. Nelson, M.~S. Pepe, K.~H. Allison, S.~J. Schnitt, et~al.
\newblock Diagnostic concordance among pathologists interpreting breast biopsy
  specimens.
\newblock {\em Jama}, 313(11):1122--1132, 2015.

\bibitem{ghaznavi2013digital}
F.~Ghaznavi, A.~Evans, A.~Madabhushi, and M.~Feldman.
\newblock Digital imaging in pathology: whole-slide imaging and beyond.
\newblock {\em Annual Review of Pathology: Mechanisms of Disease}, 8:331--359,
  2013.

\bibitem{gurcan2009histopathological}
M.~N. Gurcan, L.~E. Boucheron, A.~Can, A.~Madabhushi, N.~M. Rajpoot, and
  B.~Yener.
\newblock Histopathological image analysis: a review.
\newblock {\em Biomedical Engineering, IEEE Reviews in}, 2:147--171, 2009.

\bibitem{irshad2014methods}
H.~Irshad, A.~Veillard, L.~Roux, and D.~Racoceanu.
\newblock Methods for nuclei detection, segmentation, and classification in
  digital histopathology: A review—current status and future potential.
\newblock {\em Biomedical Engineering, IEEE Reviews in}, 7:97--114, 2014.

\bibitem{jaffer2014evolution}
S.~Jaffer and I.~J. Bleiweiss.
\newblock Evolution of sentinel lymph node biopsy in breast cancer, in and out
  of vogue?
\newblock {\em Advances in anatomic pathology}, 21(6):433--442, 2014.

\bibitem{alexnet}
A.~Krizhevsky, I.~Sutskever, and G.~E. Hinton.
\newblock Imagenet classification with deep convolutional neural networks.
\newblock In F.~Pereira, C.~J.~C. Burges, L.~Bottou, and K.~Q. Weinberger,
  editors, {\em Advances in Neural Information Processing Systems 25}, pages
  1097--1105. Curran Associates, Inc., 2012.

\bibitem{lyman2005american}
G.~H. Lyman, A.~E. Giuliano, M.~R. Somerfield, A.~B. Benson, D.~C. Bodurka,
  H.~J. Burstein, A.~J. Cochran, H.~S. Cody, S.~B. Edge, S.~Galper, et~al.
\newblock American society of clinical oncology guideline recommendations for
  sentinel lymph node biopsy in early-stage breast cancer.
\newblock {\em Journal of Clinical Oncology}, 23(30):7703--7720, 2005.

\bibitem{lyman2014sentinel}
G.~H. Lyman, S.~Temin, S.~B. Edge, L.~A. Newman, R.~R. Turner, D.~L. Weaver,
  A.~B. Benson, L.~D. Bosserman, H.~J. Burstein, H.~Cody, et~al.
\newblock Sentinel lymph node biopsy for patients with early-stage breast
  cancer: American society of clinical oncology clinical practice guideline
  update.
\newblock {\em Journal of Clinical Oncology}, 32(13):1365--1383, 2014.

\bibitem{nakhleh2006error}
R.~E. Nakhleh.
\newblock Error reduction in surgical pathology.
\newblock {\em Archives of pathology \& laboratory medicine}, 130(5):630--632,
  2006.

\bibitem{otsu}
N.~Otsu.
\newblock {A} {T}hreshold {S}election {M}ethod from {G}ray-level {H}istograms.
\newblock {\em IEEE Transactions on Systems, Man and Cybernetics}, 9(1):62--66,
  1979.

\bibitem{raab2005clinical}
S.~S. Raab, D.~M. Grzybicki, J.~E. Janosky, R.~J. Zarbo, F.~A. Meier,
  C.~Jensen, and S.~J. Geyer.
\newblock Clinical impact and frequency of anatomic pathology errors in cancer
  diagnoses.
\newblock {\em Cancer}, 104(10):2205--2213, 2005.

\bibitem{russakovsky2015imagenet}
O.~Russakovsky, J.~Deng, H.~Su, J.~Krause, S.~Satheesh, S.~Ma, Z.~Huang,
  A.~Karpathy, A.~Khosla, M.~Bernstein, et~al.
\newblock Imagenet large scale visual recognition challenge.
\newblock {\em International Journal of Computer Vision}, 115(3):211--252,
  2015.

\bibitem{vgg16}
K.~Simonyan and A.~Zisserman.
\newblock Very deep convolutional networks for large-scale image recognition.
\newblock {\em CoRR}, abs/1409.1556, 2014.

\bibitem{googlenet}
C.~Szegedy, W.~Liu, Y.~Jia, P.~Sermanet, S.~Reed, D.~Anguelov, D.~Erhan,
  V.~Vanhoucke, and A.~Rabinovich.
\newblock Going deeper with convolutions.
\newblock In {\em CVPR 2015}, 2015.

\bibitem{face}
D.~Wang, C.~Otto, and A.~K. Jain.
\newblock {Face Search at Scale: 80 Million Gallery}.
\newblock 2015.

\bibitem{weaver2003comparison}
D.~L. Weaver, D.~N. Krag, E.~A. Manna, T.~Ashikaga, S.~P. Harlow, and K.~D.
  Bauer.
\newblock Comparison of pathologist-detected and automated computer-assisted
  image analysis detected sentinel lymph node micrometastases in breast cancer.
\newblock {\em Modern pathology}, 16(11):1159--1163, 2003.

\end{thebibliography}
}

\end{document}